# SAVINGS IN LOCATION MANAGEMENT COSTS LEVERAGING USER STATISTICS

E. Martin and R. Bajcsy

Department of Electrical Engineering and Computer Science
University of California, Berkeley
California, USA
emartin@eecs.berkeley.edu

## ABSTRACT

*The growth in the number of users in mobile communications networks and the rise in the traffic generated by each user, are responsible for the increasing importance of Mobility Management. Within Mobility Management, the main objective of Location Management is to enable the roaming of the user in the coverage area. In this paper, we analyze the savings in Location Management costs obtained leveraging the users' statistics, in comparison with the classical strategy. In particular, we introduce two novel algorithms to obtain the β parameters (useful terms in the calculation of location update costs for different Location Management strategies), utilizing a geographical study of relative positions of the cells within the location areas. Eventually, we discuss the influence of the different network parameters on the total Location Management costs savings for both the radio interface and the fixed network part, providing useful guidelines for the optimum design of the networks.*

## KEYWORDS

*Mobility Management, Location Management, User statistics, Mobile Communications Networks.*

## 1. INTRODUCTION

Mobile communications networks operators need to solve difficult aspects regarding the mobility of the users and their interaction with the networks. Mobility Management is responsible for Handoff and Location Management. The former enables the continuation of a call while the user is on the move and changes cell, while the later enables the roaming of the user in the coverage area, with the main tasks involved being location update and paging [1-6]. The location update procedure consists of informing the network about every new location the mobile terminal enters, while paging is employed by the network to deliver incoming calls to the user. The signaling messages involved in these two procedures consume a significant proportion of the available radio resources [7-10]. In order to minimize this signaling burden, the location area concept (a set of cells) is used, whereby the mobile terminal will inform the network about a change in its position only when the location area's border has been crossed. The employment of the call and mobility patterns of the user can help optimize the location area's dimensions and minimize signaling costs [11]. In fact, mobile network operators often leverage handover statistics to improve the structure of their networks, with a strong impact on service performance and signaling load [12]. In this sense, user statistics-based algorithms for Location Management have proved to significantly reduce signaling costs [13-15]. In this type of algorithms, the most frequently visited location areas are assigned a probability coefficient consistent with the user's residence time in each one of them. Subsequently, the network creates a list to order the location areas according to those probabilities, and in the case of an incoming call, the location areas will be paged sequentially following their decreasing order of probability. When the mobile user exits the predetermined set of location areas, it will perform a location update operation in the first visited cell. Therefore, a profile in the form of a list is needed for each user, containing the identification of the most frequently visited location areas. In a simplified approach of this algorithm, only long term statistics (weeks or months) are memorized by the system, ignoring short term statistics (hours or days). And even this basic approach considering only long term statistics can bring important savings in location update operations. Recent examples making use of this approach can be found in reference [16], which describes an algorithm leveraging the user profile history to reduce location update costs, utilizing cascaded correlation neural networks trained on historical data of the user's movements. This approach can be further improved through the consideration of detailed data from the activity of the user, which can be extracted with the sensors embedded in current state-of-the-art smart phones [17, 18].

In this paper, we analyze the savings in Location Management costs obtained leveraging the users' statistics, in comparison with the classical strategy. In particular, we introduce two novel algorithms to obtain the β parameters (useful terms in the calculation of location update costs for different Location Management strategies), utilizing a geographical study of relative positions of the cells within the location areas. Additionally, we discuss the influence of the different network parameters on the total Location Management costs savings for both the radio interface and the fixed network part, providing useful guidelines for the optimum design of the networks.

The rest of this paper is organized as follows. In Section 2, we focus on the analysis of the location update costs for the user statistics-based algorithm and the classical strategy, making use of two new algorithms for the calculation of the β parameters. In Section 3, we analyze the paging costs for the user statistics-based algorithm and the classical strategy, while Section 4 is devoted to the study of the costs derived from maintaining the list of location areas managed by the user statistics-based algorithm. In Section 5, we examine the total Location Management costs savings of the user statistics-based algorithm in comparison with the classical strategy for both the radio interface and the fixed network parts. Conclusions are drawn in Section 6.

## 2. CALCULATION OF LOCATION UPDATE COSTS

Assuming that a user of a mobile communications network follows a random movement and that all the location areas under study have the same area, the frequency of the location updates will depend on the speed of the mobile user [19-26], $v$, and the surface and perimeter length of the location areas [27-32]. Taking into account that the location update operations can take place within a same VLR (case 1, with probability $\beta_1$), or between two VLRs, making use of the Temporary Mobile Subscriber Identity (case 2.1, with probability $\beta_{21}$), or making use of the International Mobile Subscriber Identity (case 2.2, with probability $\beta_{22}$), the location update costs for the classical strategy in mobile communications networks with a two-tier architecture can be expressed as follows [7-9]:

$$Cost\_update\_CS = \frac{8v}{\pi R \sqrt{N}} \cdot \left[ \beta_1 \cdot Nbl_{\cos, case1}(i) + \beta_{21} \cdot Nbl_{\cos, case21}(i) + \beta_{22} \cdot Nbl_{\cos, case22}(i) \right] \quad (1)$$

Where $R$ is the hexagonal cell side, $N$ is the number of cells per location area, and $Nbl_{\cos, case}(i)$ is the number of bytes generated by a location update at interface i for any of the three different cases explained before. Defining a parameter called $\beta_2$ as the probability of location update using different VLRs, $\beta_{21}$ can be approximated by 80% of $\beta_2$ [33], and $\beta_{22}$ by 20% of $\beta_2$. In Section 2.1, we will introduce two new algorithms for the calculation of these parameters.

For a typical user statistics-based algorithm, also called "Alternative Strategy (AS)" by some authors [7-8], the location update costs can be expressed as follows:

$$Cost\_update\_AS = \left(1 - \sum_{i=1}^{k} \alpha_i\right) \cdot Cost\_update\_CS \qquad (2)$$

Where $\alpha_i$ is the probability of finding a mobile user in the location area $a_i$, and $k$ is the number of location areas administered by this strategy.

## 2.1. Determination of the β parameters

Assuming densely populated areas, with an average number of cells per location area of 10 [34], and an average number of location areas managed by a VLR of 5, the calculation of the β parameters to obtain the location update costs will be tackled next.

Different algorithms can be used to obtain the values for the β parameters. In this paper we analyze the cells in the network one by one and determine the probabilities of a mobile terminal with random movement entering a new location area, whether within the same VLR or not, so that each cell is assigned a set of values, marked with a cross (denoted by "x") or a dot (denoted by "•") in Figure 1, to reflect respectively the probabilities of crossing the location area border and moving outside the actual VLR administered zone or remaining within it. This approach can be enhanced further leveraging the information obtainable from the sensors embedded in smart phones about the users' location [5, 10, 15, 25], velocity [19, 21-24, 26], and activity [17-18].

The x and • numbers could be obtained through the mobile terminal's mobility parameters owned by the network operator, or through a geographical study of relative positions of the cells within the different location areas and the VLR administered zone itself. Considering this last option, the different numbers assigned to each cell can be made dependant upon the designer's criteria, for instance in the two following ways: first, if the designer just wants to reflect the fact that a cell is neighboring a different VLR administered zone/location area, or second, if the designer wants to reflect the exact proportionality between the number of neighboring cells from a different VLR administered zone and the number of neighboring cells from different location areas within the same VLR administered zone. These two alternatives lead to a couple of methods that we respectively name simple and advanced algorithms.

### 2.1.1. Simple Algorithm

Taking for example a squared geographical area of dimensions 7·7=49 cells, so that the cells administered by a VLR can be grouped in 5 location areas with 10 cells each but one of them with 9, considering that every cell in the border of the VLR administered zone as a whole can be assigned an x, and every cell sharing border with another location area within the same zone can be assigned a •, the proportion between the number of •s and the sum of the number of xs and •s will represent the $\beta_1$ parameter, while the proportion between the number of xs and the sum of the number of xs and •s will represent the $\beta_2$ parameter. The results obtained for the referred deployment are: $\beta_1 = 40/(40+24) = 0.625$, and $\beta_2 = 24/(40+24) = 0.375$.

Considering now the same VLR administered area but with lower number of cells per location area (9,7,6), so that the number of location areas increases to 6, the results obtained are very similar: $\beta_1=41/(41+24)=0.63$ and $\beta_2=24/(41+24)=0.37$. Now taking a VLR area composed of 7·7 hexagonal cells, with 5 location areas of 11, 10 and 9 cells, the results obtained are: $\beta_1=34/(34+24)=0.59$ and $\beta_2=24/(34+24)=0.41$, similar to the previous case, although $\beta_2$ becomes noticeably larger.

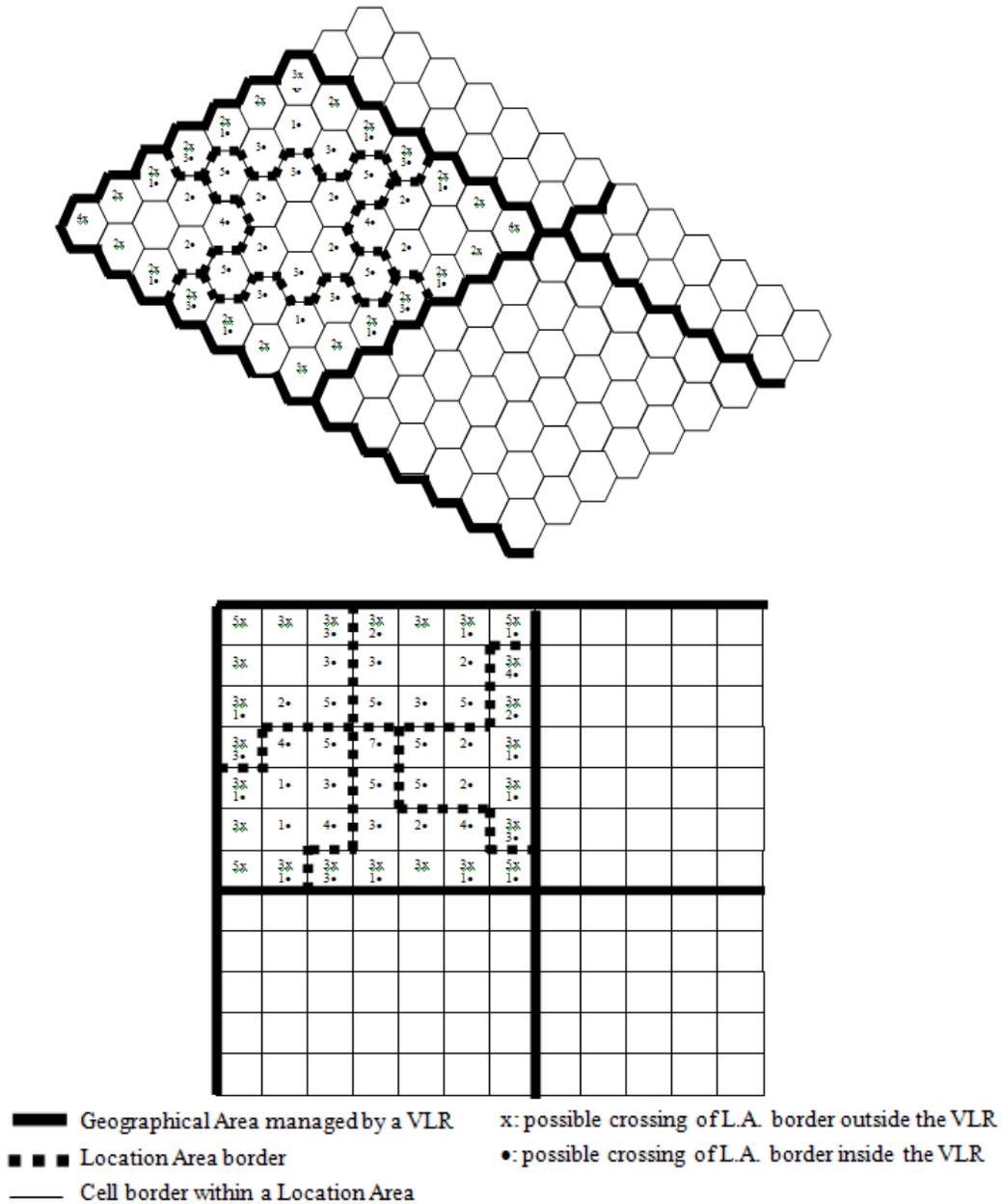

Figure 1. Calculation of basic parameters to obtain Location Management Costs, considering hexagonal or square shaped cells.

### 2.1.2. Advanced Algorithm

Taking into account for each particular cell the exact number of neighboring cells sharing location area borders whether or not being administered by the same VLR, the number of xs and •s obtained in this way rises in comparison with the simple algorithm, but the results remain quite similar for some of the cases. In this sense, for the structure of the square cells with 5 location areas per VLR, the results obtained are: $\beta_1 = 110/(110+80) = 0.58$, and $\beta_2 = 80/(110+80) = 0.42$.

For the 7·7 hexagonal cells structure, the outcome is: $\beta_1=84/(84+54)=0.61$ and $\beta_2=54/(84+54)=0.39$, again similar to previous results, although this time $\beta_2$ becomes noticeably smaller. More results obtained by means of this algorithm are presented in Table 1, and some of the geographical configurations are shown in Figure 2.

From Table 1, it can be noticed that for a same VLR administered zone dimension and cell shape, as the size of the location areas rises, $\beta_1$ declines and complementarily, $\beta_2$ grows. Regarding the number of xs, it remains constant regardless of the location areas shape and size for a fixed geographical area covered by the VLR, as this number just depends on the size and shape of that VLR area. In order to minimize the number of xs in proportion to •s, and therefore decrease the values of the $\beta_{21}$ and $\beta_{22}$ parameters, the VLR area should be as regular as possible, and containing the largest possible number of cells within (for instance, 100 hexagonal cells served by an only VLR bring 78 xs, while two groups of 49 square cells served by an VLR each, bring 108 xs). Furthermore, considering a VLR area of $m·m$ cells, the number of xs in a square cells deployment will be 20+12·($m$-2), while for hexagonal cells, the number of xs will be 14+8·($m$-2), considerably lower.

The number of •s depends on the size and shape of the location areas. The smaller the location areas, the larger the total length of shared borders and, consequently the larger the number of •s. In the same sense, the more irregular the shape of the location areas, the larger the number of •s. Obviously, for a fixed size of location areas, the larger the geographical zone covered by the VLR, the higher the number of •s. In order to minimize the number of •s, and therefore diminish the values of the $\beta_1$ parameter, the shape of the location areas should be square, and their size as large as possible, ideally to fit one location area in one VLR zone.

Making use of the calculated β parameters, the location update costs for different cellular deployments and Location Management strategies will be obtained next.

## 2.2. Analysis of the Location Update costs for the user statistics-based Algorithm and the Classical Strategy

Analyzing multiple network deployments administered by the user statistics-based algorithm, with varying numbers of location areas in the list, and different sets of probabilities for the location areas, we can determine the evolution of the location update costs with each particular scheme, making use of the previously introduced algorithms for the calculation of the β parameters. A representative sample of these results is shown in Figure 3, where we can observe that the location update costs are minimized when the user statistics based algorithm is employed, and this minimization is proportional to the summation of the probabilities of the location areas managed by the user statistics-based algorithm, regardless of the actual number of those location areas. From the study of multiple configurations, we can conclude that the larger the VLR administered zone, the higher the decreasing speed of the location update costs with the number of cells per location area and, also, for the same size of the deployment structure, the speed of the descent is higher for the hexagonal cells structure than for the square cells one.

Table 1. Calculation of β parameters for different network deployments.

| Cell Shape | VLR administered zone dimension | Number of L.A.s per VLR | Number of cells per L.A. | Regularity of shape of L.A.s | No. x | No. ● | $\beta_1$ | $\beta_2$ | $\beta_{21}$ | $\beta_{22}$ |
|---|---|---|---|---|---|---|---|---|---|---|
| Hexagonal | 7cells·7cells | 5 | 9,10,11 | Good | 54 | 84 | 0.61 | 0.39 | 0.312 | 0.078 |
| Hexagonal | 7cells·7cells | 4 | 9,12,16 | Very good | 54 | 50 | 0.48 | 0.52 | 0.416 | 0.104 |
| Hexagonal | 10cells·10cells | 9 | 9,12,16 | Very good | 78 | 144 | 0.65 | 0.35 | 0.28 | 0.07 |
| Hexagonal | 10cells·10cells | 4 | 25 | Very good | 78 | 74 | 0.49 | 0.51 | 0.408 | 0.102 |
| Hexagonal | 10cells·10cells | 2 | 50 | Very good | 78 | 38 | 0.33 | 0.67 | 0.536 | 0.134 |
| Square | 7cells·7cells | 17 | 2,3 | Good | 80 | 248 | 0.76 | 0.24 | 0.192 | 0.048 |
| Square | 7cells·7cells | 16 | 1,2,4 | Very good | 80 | 191 | 0.7 | 0.3 | 0.24 | 0.06 |
| Square | 7cells·7cells | 9 | 4,6,9 | Very good | 80 | 136 | 0.63 | 0.37 | 0.296 | 0.074 |
| Square | 7cells·7cells | 6 | 6,8,9,12 | Very good | 80 | 106 | 0.54 | 0.41 | 0.328 | 0.082 |
| Square | 7cells·7cells | 5 | 9,10 | Medium | 80 | 110 | 0.58 | 0.42 | 0.336 | 0.084 |
| Square | 7cells·7cells | 4 | 9,12,16 | Very good | 80 | 72 | 0.47 | 0.53 | 0.424 | 0.106 |
| Square | 7cells·7cells | 3 | 12,16,21 | Good | 80 | 50 | 0.38 | 0.62 | 0.496 | 0.124 |
| Square | 7cells·7cells | 2 | 21,28 | Very good | 80 | 38 | 0.32 | 0.68 | 0.544 | 0.136 |
| Square | 10cells·10cells | 33 | 3,4 | Good | 116 | 550 | 0.83 | 0.17 | 0.136 | 0.034 |
| Square | 10cells·10cells | 16 | 4,6,9 | Very good | 116 | 300 | 0.72 | 0.28 | 0.224 | 0.056 |
| Square | 10cells·10cells | 9 | 3,4,12,15 | Good | 116 | 208 | 0.64 | 0.36 | 0.288 | 0.072 |
| Square | 10cells·10cells | 9 | 9,12,16 | Very good | 116 | 208 | 0.64 | 0.36 | 0.288 | 0.072 |
| Square | 10cells·10cells | 4 | 25 | Very good | 116 | 108 | 0.48 | 0.52 | 0.416 | 0.104 |
| Square | 10cells·10cells | 3 | 30,40 | Very good | 116 | 112 | 0.49 | 0.51 | 0.408 | 0.102 |
| Square | 10cells·10cells | 2 | 50 | Very good | 116 | 56 | 0.33 | 0.67 | 0.536 | 0.134 |

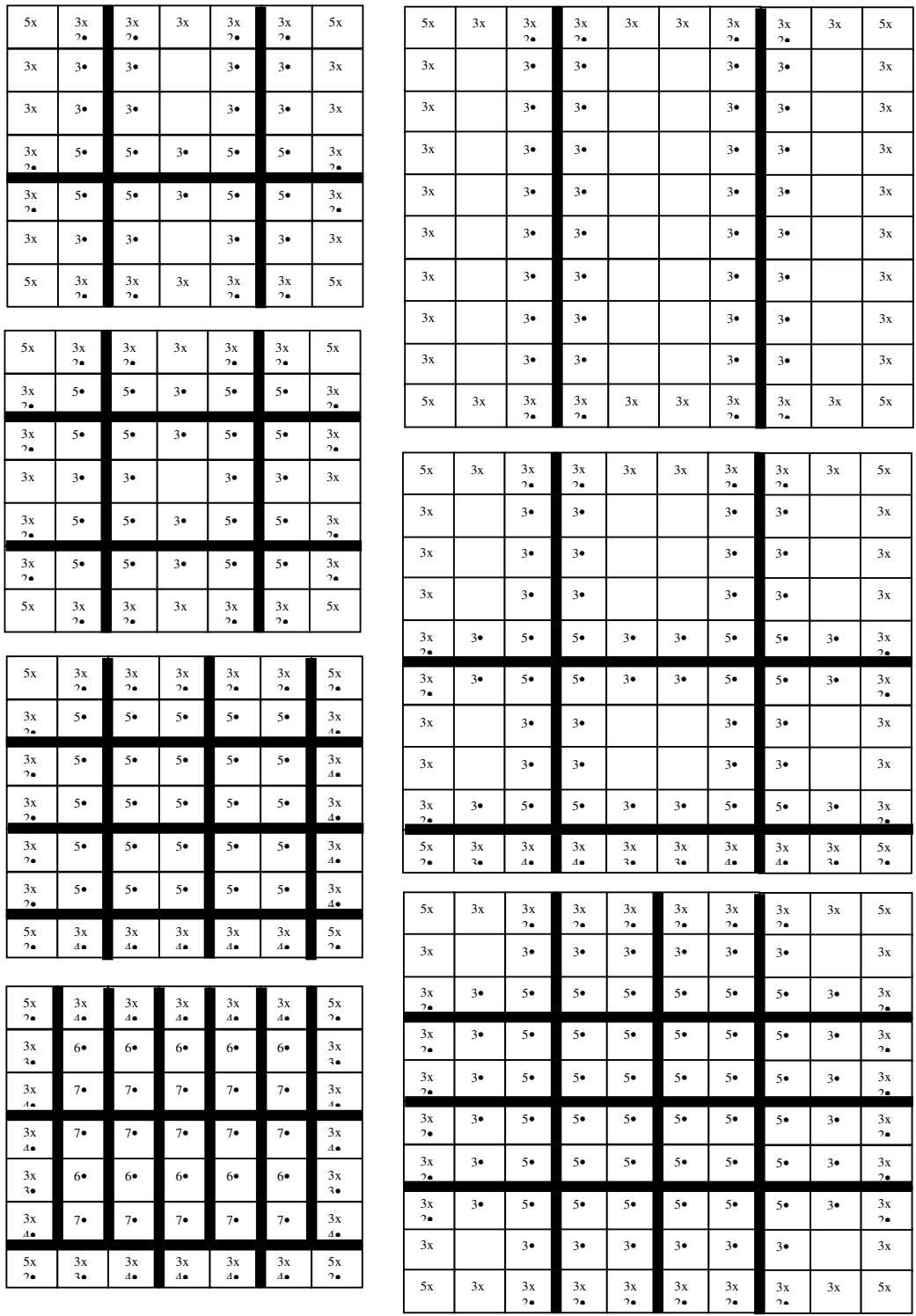

Figure 2. Examples of calculation of the β parameters for different VLR administered zones and different location area structures for square cells.

Example Sets of 3 Location Areas:

Probabilities: α1=0.4, α2=0.1, α3=0.05
Probabilities: α1=0.5, α2=0.1, α3=0.05
Probabilities: α1=0.6, α2=0.1, α3=0.05
Probabilities: α1=0.7, α2=0.1, α3=0.05
Probabilities: α1=0.8, α2=0.1, α3=0.05

Example Sets of 5 Location Areas:
Probabilities: α1=0.4, α2=0.1, α3=0.05, α4=0.02, α5=0.01
Probabilities: α1=0.5, α2=0.1, α3=0.05, α4=0.02, α5=0.01
Probabilities: α1=0.6, α2=0.1, α3=0.05, α4=0.02, α5=0.01
Probabilities: α1=0.7, α2=0.1, α3=0.05, α4=0.02, α5=0.01
Probabilities: α1=0.8, α2=0.1, α3=0.05, α4=0.02, α5=0.01

Example Sets of 9 Location Areas:
Probabilities: α1=0.4, α2=0.05, α3=0.03, α4=0.02, α5=0.01, α6=0.008, α7=0.005, α8=0.003, α9=0.002
Probabilities: α1=0.5, α2=0.05, α3=0.03, α4=0.02, α5=0.01, α6=0.008, α7=0.005, α8=0.003, α9=0.002
Probabilities: α1=0.6, α2=0.05, α3=0.03, α4=0.02, α5=0.01, α6=0.008, α7=0.005, α8=0.003, α9=0.002
Probabilities: α1=0.7, α2=0.05, α3=0.03, α4=0.02, α5=0.01, α6=0.008, α7=0.005, α8=0.003, α9=0.002
Probabilities: α1=0.8, α2=0.05, α3=0.03, α4=0.02, α5=0.01, α6=0.008, α7=0.005, α8=0.003, α9=0.002

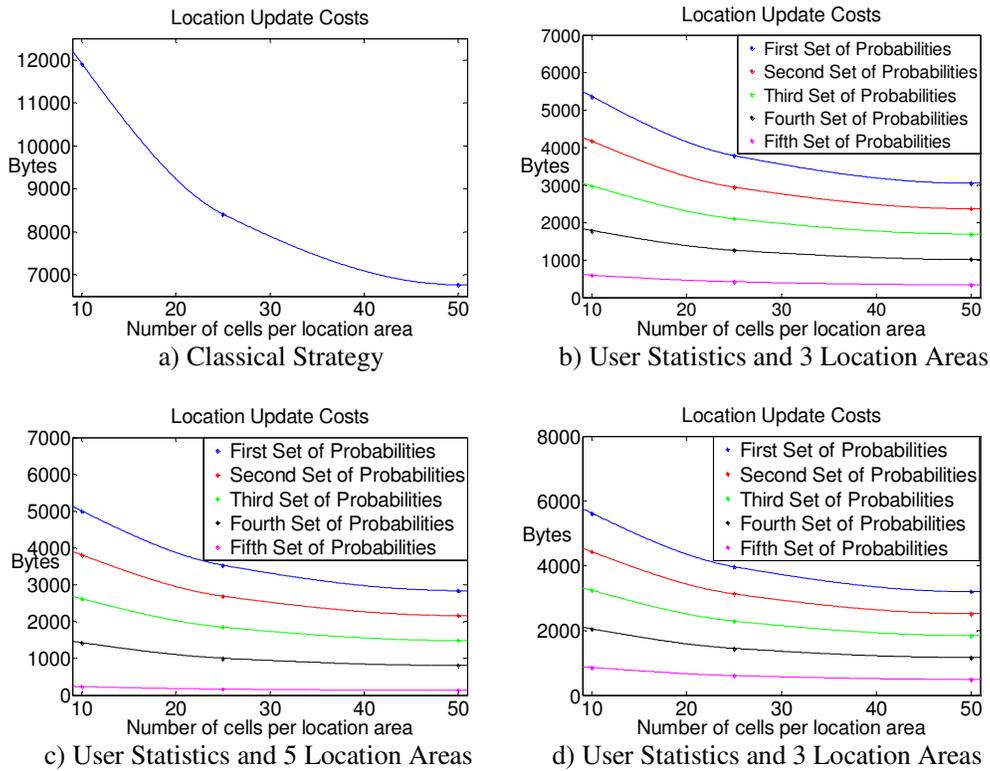

Figure 3. Example of location update costs for the classical strategy and the user statistics-based algorithm considering different numbers of Location Areas managed.

The fact that hexagonal cells deliver lower location update costs than square cells (in agreement with [2-4]) can be reasoned making use of the advanced algorithm for the calculation of the β parameters: the percentage reduction in the hexagonal cell structures with respect to the square cells structures is always higher for the number of xs than for the number of •s, as shown in Table 2.

Table 2. Comparison of the percentage reduction of xs and •s in the advanced algorithm for the hexagonal cells with respect to the square cells.

| VLR administered zone size | No. Location Areas | No. cells per L. A. | Percentage of reduction in x | Percentage of reduction in • |
|---|---|---|---|---|
| 10·10 | 9 | 11 | 32.76 | 30.77 |
| 10·10 | 4 | 25 | 32.76 | 31.48 |
| 10·10 | 2 | 50 | 32.76 | 32.14 |
| 7·7 | 5 | 10 | 32.5 | 23.63 |
| 7·7 | 4 | 11 | 32.5 | 30.55 |

Therefore, the hexagonal cells structures will present relatively lower values of $\beta_{21}$ and $\beta_{22}$, which account for the highest terms in the location update costs, and consequently the costs will be lower. However, from Table 2, it can be inferred that as the number of cells per location area increases, the difference in the percentage reduction between xs and •s tends to decline, and consequently the reduction in the location update costs will diminish.

## 2.2. Study of the Location Update costs focusing on the Rate of Updates

Considering a fluid mobility model, the number of mobile terminals exiting an area in a time unit is proportional to the product of terminals density, their average velocity and the length of the area perimeter [35], and all these parameters can be integrated within the concept of "rate of location updates" [2-4]. Taking average values for the mobility parameters from [36], the behavior of the rate of location updates with the number of cells per location area will be analyzed next.

$$Rate\_{update} = \frac{velocity \cdot Perimeter}{Number\_cells \cdot \pi \cdot Surface\_cell} \quad (3)$$

In (3), the location area perimeter can be substituted by $L \cdot \sqrt{Number\_cells}$, where $L$ is the average cell perimeter [37]. In a similar way, the rate of location updates for a mobile user inside the location areas in the list managed by the user statistics-based algorithm can be studied dividing (3) by the square root of the number of location areas in the list.

As observed in Figure 4, the larger the cell size, the lower the update rates. Besides, the larger the number of location areas in the list, the lower the rates, although it can be noticed that the drops in the rates become proportionally smaller as the number of location areas in the list go above 8 location areas. Consequently, in terms of location update costs savings, implementations of the user statistics-based algorithm with more than 8 location areas could risk a loss in efficiency if the increased paging and list maintaining costs are taken into account. In the same sense, the reductions in the rate of updates when the number of cells per location area exceeds 15 are very small for all the ranges of numbers of location areas in the list and cell sides analyzed.

Regarding (3), it is interesting to note that until recently, the estimation of the user velocity represented a difficult task, especially for indoor environments in which GPS sensors do not work. This issue can now be easily solved leveraging the WiFi radio and the accelerometer and magnetometer sensors embedded in current state-of-the-art smart phones [5, 10, 15, 19, 21-26].

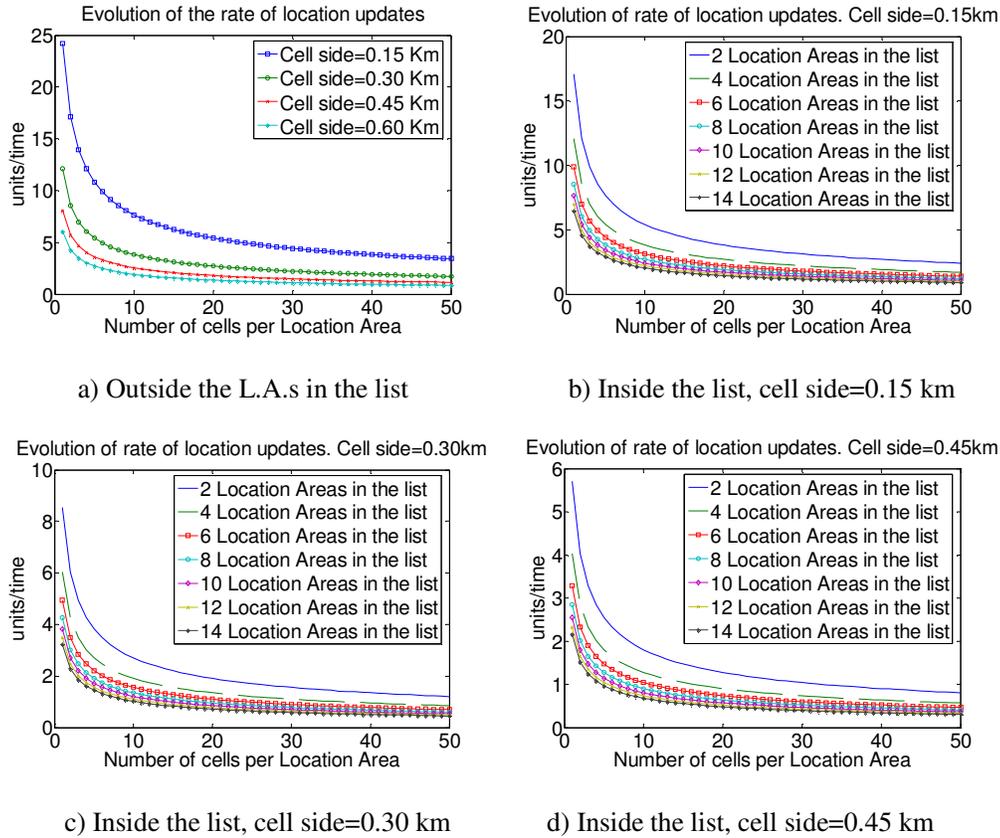

Figure 4. Rate of location updates outside and inside the location areas in the list.

## 3. ANALYSIS OF THE PAGING COSTS

### 3.1. Calculation of the Paging costs for the Classical Strategy

The paging costs in the classical strategy can be expressed as follows:

$$Cost\_paging\_CS = N \cdot [\lambda_{t1} \cdot Nbp_{\cos 1}(i) + \lambda_{t2} \cdot Nbp_{\cos 2}(i)] \qquad (4)$$

Where $\lambda_{t1}$ is the rate of mobile terminating calls, $\lambda_{t2}$ is the rate of unsuccessful attempts, $Nbp_{\cos}(i)$ represents the number of bytes generated by a paging process at interface i in case 1 (successful attempt) or case 2 (unsuccessful attempt), and $N$ is the number of cells per location area. Considering typical values of $\lambda_{t1}$=0.6 calls/hour, $\lambda_{t2}=\lambda_{t1}/100$, and the number of bytes generated at the radio interface as 21.5 for case 1 and 83 for case 2 [38], the paging costs can be expressed as $Cost\_{paging\_CS}$ = 13.39·$N$. Next, we will compare these costs with those obtained by means of the user statistics-based algorithm.

### 3.2. Calculation of the Paging costs for the user statistics-based Algorithm

For the user statistics-based algorithm, the paging costs can be expressed as follows [38]:

$$Cost\_paging\_AS = \{ (\sum_{i=1}^{k}\alpha_i \cdot [\sum_{i=1}^{k}[\alpha_i \cdot (\lambda_{t1} \cdot Nbp_{\cos 1}(i) + \lambda_{t2} \cdot Nbp_{\cos 2}(i)) +$$
$$+ (1-\alpha_i) \cdot Nbp_{\cos 2}(i) \cdot (\lambda_{t1} + \lambda_{t2}) \cdot (1 - \sum_{j=1}^{i-1}\alpha_j) \cdot (1 - \frac{\lambda_{t2}}{\lambda_{t1}+\lambda_{t2}})]) + \quad (5)$$
$$+ [1 - \sum_{i=1}^{k}\alpha_i] \cdot (\lambda_{t1} \cdot Nbp_{\cos 1}(i) + \lambda_{t2} \cdot Nbp_{\cos 2}(i))) \} \cdot N$$

Plotting these costs for different location areas probabilities, we obtain Figure 5.

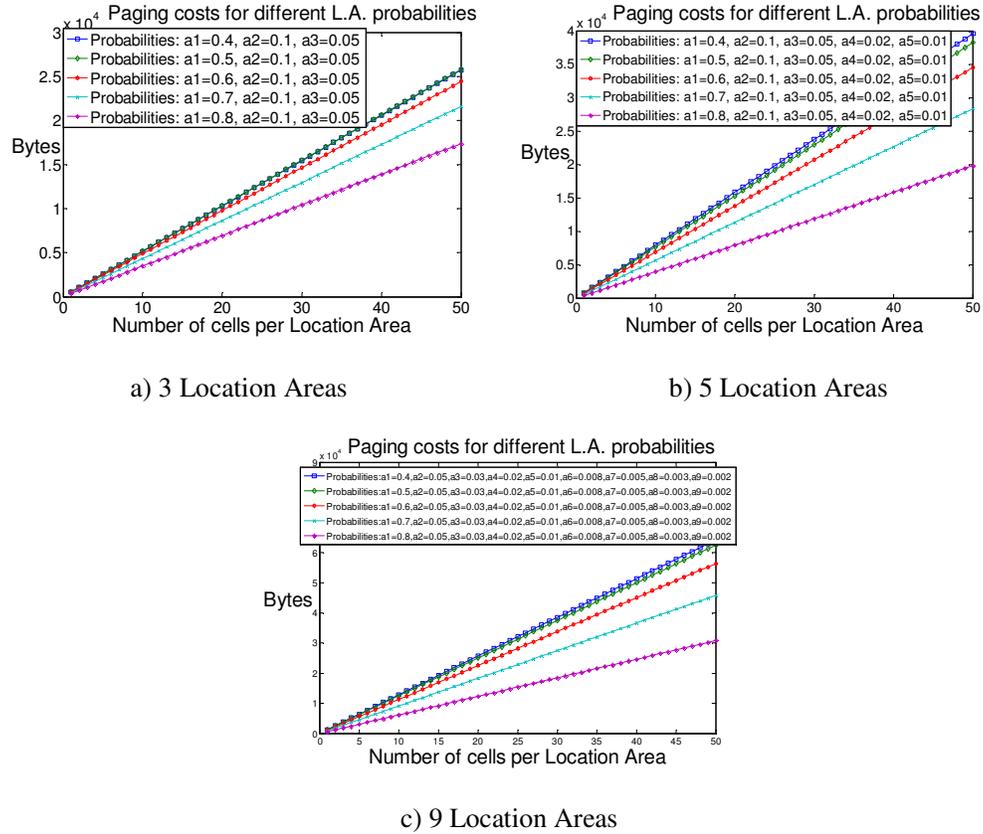

a) 3 Location Areas  b) 5 Location Areas

c) 9 Location Areas

Figure 5. Variation of the paging costs for the user statistics-based algorithm with different numbers of location areas and different probability sets.

In agreement with a general model of Location Management costs [11], the paging costs present a linear behavior with the number of cells in the location areas, and for a fixed number of location areas managed by the user statistics-based algorithm, the higher the summation of the probabilities of the location areas, the lower the paging costs, as observed in Figure 5. It can also be noticed that the reductions in the paging costs slope due to rises in the probability of the first location area, are more important as the summation of the probabilities of all the location areas managed approaches unity, and also as the number of those location areas increases (e.g. measured fall of 53.8% in the paging costs slope as $\alpha_1$ goes from 0.4 to 0.8 for 9 location areas, while for 3 location areas, the same rise in $\alpha_1$ brings a drop in the slope of only 34.6%). Coherently with expected results shown in recent research works in this field[2-4], measurements show that if the rate of mobile terminated calls is decreased/increased (the rate of unsuccessful call attempts is decreased/increased proportionally), the paging costs decrease/increase, and the

International Journal of UbiComp (IJU), Vol.2, No.3, July 2011

previously described behavior of the slopes, related to the location areas probabilities, is maintained. In comparison with the classical strategy, the paging costs for the user statistics-based algorithm for all the cases analyzed are higher; and the larger the number of cells per location area and the number of location areas managed, the higher the difference. Specifically, considering the best performance cases in the user statistics-based algorithm for each one of the location areas sets, the angle of the paging costs line with the abscissa axis rises as the number of managed location areas grows, and for 3, 5 and 9 location areas, this angle is respectively 89.83°, 89.85° and 89.90°, while for the classical strategy the referred angle remains constant at 85.72°.

### 3.3. Analysis of the Paging costs focusing on the Rate of Paging

Interesting results regarding the paging costs can also be inferred focusing on the "rate of paging". Specifically, the normalized cost of paging can be expressed as follows:

$$Cost\_normalized\_paging = Cost\_paging\_area \cdot Rate\_paging \cdot [1 + P_{inside} \cdot Cost\_next\_paging \cdot (E[N]-1)] \qquad (6)$$

Where $Cost\_paging\_area$ is the cost to page a mobile user in a single location area, $Rate\_paging$ is the rate of paging per user, $Cost\_next\_paging$ is the cost fraction of the first paging attempt corresponding to subsequent paging attempts, and E[N] is the expected number of location areas where the mobile user will be paged.

Assuming the cost of paging a user in a single location area approximately 1/17 of the cost to update its location [39], and considering $Rate\_paging$ being independent of the mobility of the user, Figures 6 and 7 show the behavior of the paging costs with the expected number of location areas paged. For this purpose, the probability of the mobile terminal being within the list of location areas is varied between 0.2 and 1, the cost fraction of subsequent paging attempts (F in the figures) is varied between 0.2 and 1.5 to analyze both cases in which subsequent attempts are respectively less or more costly.

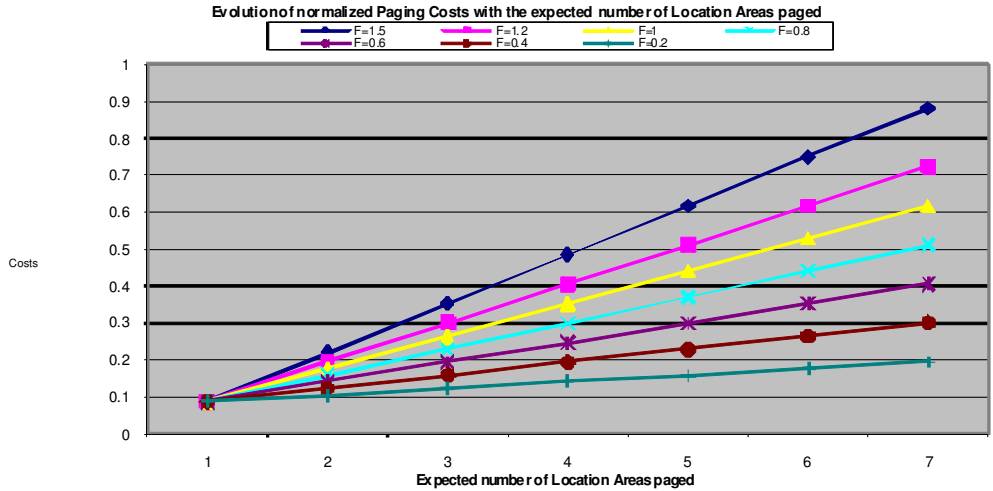

Figure 6. Evolution of the normalized paging costs with the expected number of Location Areas paged with the probability of being within the list = 1

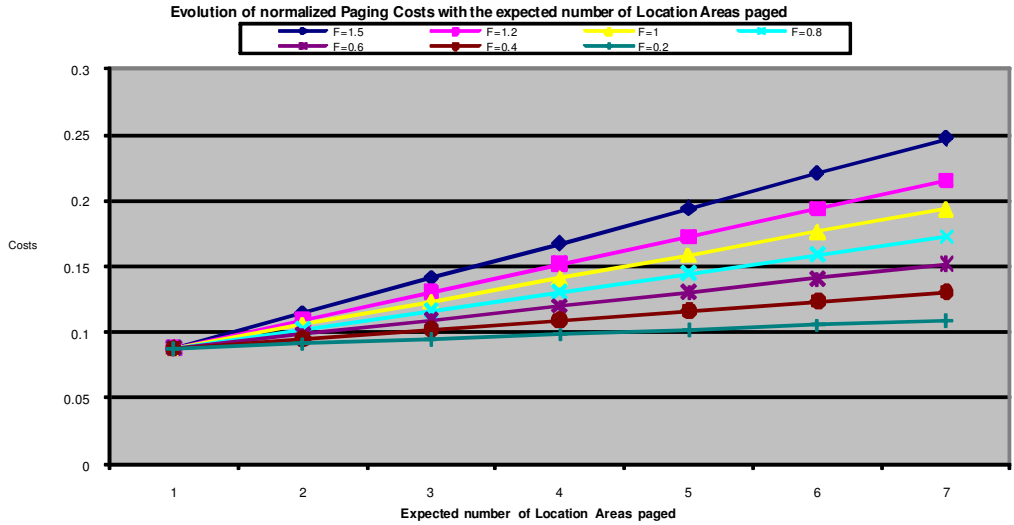

Figure 7. Evolution of the normalized paging costs with the expected number of Location Areas paged with the probability of being within the list = 0.2

As observed in Figures 6 and 7, the paging costs follow a straight linear behavior with positive slope over the expected number of location areas paged, and the slope increases with the values of $Cost\_next\_paging$ and $P_{inside}$.

## 4. COST OF LIST MAINTENANCE

Apart from the location update and paging costs, the list of administered location areas must be updated, implying a cost, which will depend on the number of location areas in the list. The normalized cost of list maintenance can be expressed as follows:

$$Cost\_normalized\_list = \frac{\lambda_{mobile\_term} + \lambda_{mobile\_orig}}{N_{umber\_calls\_to\_update}} \cdot Cost\_list \qquad (7)$$

Where $\lambda_{mobile\_term}$ is the mobile terminated calls rate, $\lambda_{mobile\_orig}$ is the mobile originated calls rate, $N_{umber\_calls\_to\_update}$ is the number of calls set by the system to update the list, and $Cost\_list$ is the cost for a single update of the list.

## 5. LOCATION MANAGEMENT COSTS SAVINGS OF THE USER STATISTICS-BASED ALGORITHM COMPARED TO THE CLASSICAL STRATEGY

Taking into account that the total cost of the user statistics-based algorithm will be the sum of the location update, paging and list maintenance costs, the savings in comparison with the classical strategy, for which the total costs were

$Cost\_update \cdot Rate\_update + Cost\_paging\_area \cdot Rate\_paging$, can be expressed as follows:

$$Savings = Pinside \cdot \left[ Cost\_update \cdot Rate\_update \cdot \left(1 - \frac{1}{\sqrt{N}}\right) - \right.$$
$$Cost\_paging\_area \cdot Rate\_paging \cdot Cost\_next\_paging \cdot (E[N]-1) \Big] - \quad (8)$$

$$\frac{\lambda_{mobile\_term} + \lambda_{mobile\_orig}}{N_{umber\_calls\_to\_update}} \cdot Cost\_list$$

By means of (8), the performance of the user statistics-based algorithm can be evaluated in both the radio interface and the fixed network parts.

## 5.1. Location Management costs savings in the radio interface

Since the list maintenance costs have no influence in the radio interface (this procedure is managed over the fixed network part), the calculation of the costs savings brought by the user statistics-based algorithm is simplified.

If we consider a uniform distribution for the user's location probabilities (yielding the poorest results in terms of costs savings, and providing us with a lower bound in performance), we obtain the results in Figure 8.

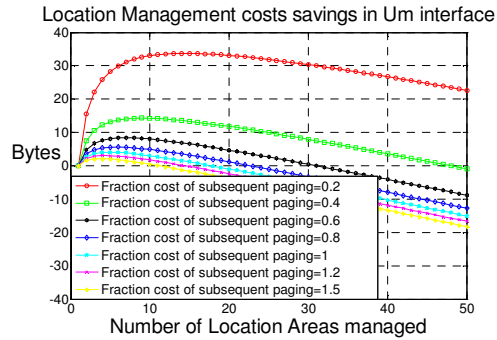

Figure 8. Location Management costs savings in the radio interface with a uniform distribution for the user's location probabilities, with $Cost\_update / Cost\_paging\_area = 17$ and $Rate\_paging / Rate\_update = 1.549$

From Figure 8, it can be observed that the larger the fraction costs of subsequent paging attempts, the lower the costs savings achieved. In other words, the higher the fraction costs of subsequent paging, the lower the number of location areas that the user statistics-based algorithm can successfully manage in order to achieve costs savings in the radio interface. Typical values of the fraction costs allow managing over 20 location areas and still achieve savings, and for example, for a fraction cost of 0.2, the number of location areas that could be managed with savings would reach 100.

The optimum number of location areas managed to obtain the largest costs savings in the radio interface ranges from 4 to 15 for fraction costs varying from 1.5 to 0.2 respectively. It must be noticed that rises in the fraction costs above 1 produce very small reductions in the optimum number of location areas (from 5 to 4 for fraction costs increasing from 1 to 1.5), while slight drops in the fraction costs below 1 bring considerable rises in the optimum number of location areas (from 6 to 15 for fraction costs falling from 0.8 to 0.2) and the costs savings themselves (7 times growth when the fraction costs diminish from 0.8 to 0.2).

For common values of the fraction cost of subsequent paging attempts (0.8) and the quotient between the single update and paging costs (17), the ratio between the paging rate and the update rate can provide us with insights into the evolution of the Location Management costs savings with the number of cells in the location areas, the cell side, the call arrival rate and the mobile user speed.

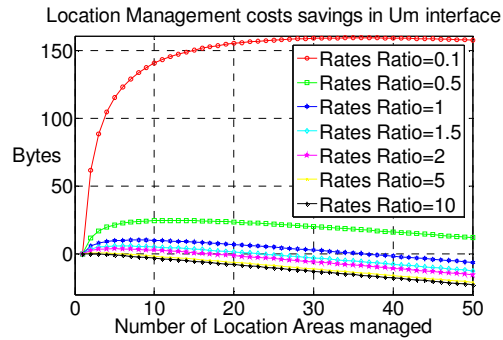

Figure 9. Location Management costs savings in radio interface for the uniform distribution, for fraction cost of subsequent paging = 0.8 and $Cost\_update / Cost\_paging\_area = 17$ and varying the $Rate\_paging / Rate\_update$ ratio.

As observed in Figure 9, when the ratio between the rates of paging and update increases (because of growths in the number of cells per location area, the cell side and the call arrival rate, or decreases in the mobile user's speed), the costs savings in the radio interface diminish and the amount of location areas that can be managed by the user statistics-based algorithm achieving savings also decreases. Specifically, for rates ratios above 10, no savings at all are achieved with the user statistics-based algorithm. On the other hand, reductions in the ratio of rates yield important increases in the costs savings, especially for values below 0.5; for instance, the optimum number of location areas goes above 13 for a ratio of 0.5, and above 40 for a ratio of 0.1. Consequently, to achieve the best results in terms of Location Management costs savings in the radio interface with the user statistics-based algorithm, on the one hand, the size of the location areas managed should be reduced by means of either decreasing the number of cells per location area or reducing the cell side, or on the other hand, the strategy should be applied to mobile users with low call arrival rates and high speed (i.e. low call-to-mobility ratios).

## 5.2. Location Management costs savings in the fixed network part

The number of location areas in the list that optimizes the costs savings in the fixed network part (obtained through derivative of (8)) is proportional to the product of $N_{umber\_calls\_to\_update}$ and $P_{inside}$. Consequently, if the frequency of updating the list is low ($N_{umber\_calls\_to\_update}$ rises) or if the mobile user is prone to stay most of the time within the location areas in the list ($P_{inside}$ grows by enlargements of the size of the location areas in the list or increases in their number or reductions in the random mobility of the user), the optimum number of location areas in the list will rise. However, recalling (8), and assuming $Cost\_list$ proportional to the number of location areas in the list, that number of location areas has to be constrained in order to obtain savings in the fixed network part. Nevertheless, the list maintenance costs are generally lower than the location update and paging costs, and the fixed network part does not represent an important constraint for the practical implementation of the user statistics-based algorithm.

Specifically, considering $Rc$ as the ratio between the cost of a single location update and the cost of a single list update, the evolution of the fixed network costs savings (assuming the cost of list maintenance to be proportional to the number of location areas in the list) can be observed in Figure 10.

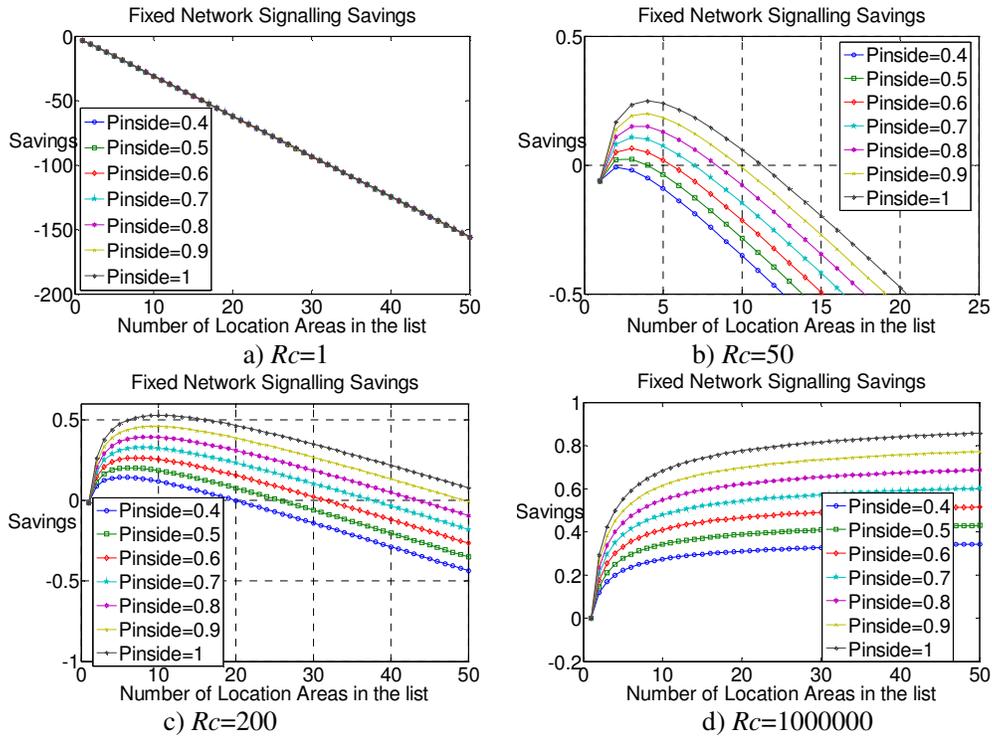

Figure 10. Fixed Network Savings with Cost of List Maintenance proportional to the number of Location Areas in the list, varying *Pinside* and the Ratio of costs between location update and list maintenance.

It is interesting to note that only when *Rc* rises above 30, some positive savings can be obtained in the fixed network part for very high values of *Pinside* ($> 0.5$), and very low numbers of location areas in the list ($< 5$).

It must also be noticed that as *Rc* rises, the savings in the fixed network part increase, and these savings present a decline in the falling speed with the number of location areas in the list above the optimum. In fact, for very high values of *Rc*, theoretically canceling the list maintenance cost, the referred slope becomes positive for all values of the number of location areas in the list.

## 6. CONCLUSIONS

In this paper, we have analyzed the Location Management costs of the user statistics-based algorithm, and compared them with the costs brought by the classical strategy. From the application of the novel algorithms proposed in Section 2.1 to obtain the β parameters (useful to calculate the location update costs for different Location Management strategies), we draw the following conclusions:

- The minimization of the $β_1$ parameter is achieved through enlargements in the location area size, ideally with square shape and fitting in the surface of a VLR administered zone.

- The minimization of the $\beta_{21}$ and $\beta_{22}$ parameters requires reductions in the size of the location areas and rises in the number of cells within the VLR administered zone, whose shape should be as regular as possible.

From the analysis of the location update costs for the user statistics-based algorithm, we can infer the following guidelines:

- Increases in the VLR administered zone size (keeping the number of cells per location area fixed) bring declines in the location update costs and rises in their decreasing speed with the number of cells per location area.

- Hexagonal cells schemes deliver lower location update costs and higher decreasing speeds in those costs than the square ones, although the difference is reduced as the number of cells per location area grows.

- The larger the summation of the probabilities controlled by the user statistics-based algorithm, the lower the location update costs, regardless of the actual number of those location areas.

Regarding the paging costs in the user statistics-based algorithm, we can observe the following trends:

- The paging costs decline as the summation of the probabilities of the location areas administered by the user statistics-based algorithm grows.

- The slope of the paging costs grows with the value of the fraction costs for subsequent paging attempts ($Cost\_next\_paging$) and the probability of the user being within the location areas in the list ($Pinside$).

In connection with the Location Management costs savings of the user statistics-based algorithm for the radio interface, we can highlight the following interesting points:

- The optimum number of location areas to keep in the list ranges from 4 to 15 for values of $Cost\_next\_paging$ varying from 1.5 to 0.2 respectively.

- Increases in $Cost\_next\_paging$ above unity produce very small declines in the optimum number of location areas (e.g. from 5 to 4 as $Cost\_next\_paging$ grows from 1 to 1.5), while reductions in $Cost\_next\_paging$ under 1 deliver noticeable rises in the optimum number of location areas (e.g. from 6 to 15 when $Cost\_next\_paging$ drops from 0.8 to 0.2) and the costs savings themselves (e.g. 7 times growth as $Cost\_next\_paging$ falls from 0.8 to 0.2).

- The performance of the user statistics-based algorithm in comparison with the classical strategy improves as the users' call-to-mobility ratios drop (preferably below 0.5).

And for the Location Management costs savings in the fixed network part, assuming the cost of a single update of the list ($Cost\_list$) to be proportional to the number of location areas in the list, we can draw the following conclusions:

- For values of the ratio between the costs of a single location update and a single list update (*Rc*) underneath 30, no savings are obtained in the fixed network part, and the larger the number of location areas administered, the larger the losses.

- For a particular value of *Rc*, the lower *Pinside*, the bigger the losses or the smaller the savings, depending on the particular value of *Rc*.

- The optimum number of location areas in the list rises with *Pinside*, and this growth is emphasized with increases in *Rc*.

- For the highest achievable values of *Rc* (over 50), the optimum number of location areas in the list to obtain savings in the fixed network part ranges between 3 and 10, depending on *Pinside* (preferably above 0.4).

In conclusion, the user statistics-based algorithm outperforms the classical strategy, especially for the highest values (> 0.5) of the mobility predictability level of the location areas most frequently visited by the user. And in order to optimize its performance, the most favorable number of location areas to maintain in the list would range from 4 to 8, keeping the number of cells per location area for densely populated areas below 15.

**Authors**

**E. Martin** is carrying out research in the Department of Electrical Engineering and Computer Science at University of California, Berkeley. He holds a MS in Telecommunications Engineering from Spain and a PhD from England within the field of location management for mobile telecommunications networks. He has research experience in both industry and academia across Europe and USA, focusing on wireless communications, sensor networks, signal processing and localization.

**R. Bajcsy** received the Master's and Ph.D. degrees in electrical engineering from the Slovak Republic, and the Ph.D. in computer science from Stanford University, California. She is a Professor of Electrical Engineering and Computer Sciences at the University of California, Berkeley. Prior to joining Berkeley, she headed the Computer and Information Science and Engineering Directorate at the National Science Foundation. Dr. Bajcsy is a member of the National Academy of Engineering and the National Academy of Science Institute of Medicine as well as a Fellow of the Association for Computing Machinery (ACM) and the American Association for Artificial Intelligence. In 2001, she received the ACM/Association for the Advancement of Artificial Intelligence Allen Newell Award, and was named as one of the 50 most important women in science in the November 2002 issue of *Discover Magazine*.